\documentclass[aps,showpacs,onecolumn,notitlepage,floatfix,amsmath,amssymb,nofootinbib]{revtex4-1}

\usepackage{amssymb}
\usepackage{amsmath}
\usepackage{epsfig}
\usepackage{graphicx}
\usepackage{color}
\usepackage{latexsym}
\usepackage{bm,latexsym}
\usepackage[mathscr]{eucal}%\usepackage{mathrsfs}
\usepackage{float}
\usepackage{pbox}

\begin{document}

\title{ New exact traversable wormhole solution to the Einstein-scalar-Gauss-Bonnet Equations coupled to 
a power-Maxwell electrodynamics }

\author{Pedro Ca\~nate$^{1}$}
\email{pcanate@fis.cinvestav.mx, pcannate@gmail.com}

\author{Nora Breton$^{1}$}
\email{nora@fis.cinvestav.mx}

\affiliation{$^{1}$Departamento de F\'isica, Centro de Investigaci\'on y de Estudios Avanzados
del I.P.N.,\\ Apdo. 14-740, Mexico City, Mexico.\\
}

\begin{abstract} 
We present a novel, exact, traversable wormhole (T-WH) solution for $(3+1)$-dimensional Einstein-scalar-Gauss-Bonnet theory (EsGB) coupled  to a power-Maxwell nonlinear electrodynamics (NLED). The solution is characterized by two parameters, $\mathcal{Q}\!_{\rm e}$ and $\mathcal{Q}\!_{_{ \mathcal{S} }}$, associated respectively with the electromagnetic field and the scalar field. We show that for $\mathcal{Q}^2_{\rm e}  - \mathcal{Q}\!_{_{ \mathcal{S} }}>0$ the solution can be interpreted as a  traversable wormhole. In the  general case, with non-vanishing electromagnetic field, the scalar-Gauss-Bonnet term (sGB) is the only responsible for the negative energy density necessary for the traversability. In the limiting case of vanishing electromagnetic field, the scalar field becomes a phantom one keeping the WH throat open and in this case the Ellis WH solution \cite{Ellis} is recovered. 
\end{abstract}

%
% 04.20.Jb Exact solutions
% 04.50.Kd Modified theories of gravity
% 04.50.−h Higher-dimensional gravity and other theories of gravity
% 04.60.Cf Gravitational aspects of string theory
\maketitle

%{\bf Introduction.} 
\section{Introduction}

Traversable Wormholes (T-WHs) of the Morris \& Thorne type \cite{ThorneMorris} are fascinating predictions of General Relativity (GR), which can be used as interstellar travel shortcuts between two distant regions of the universe. In GR the construction of T-WHs necessarily demands the presence of a type of matter, known as exotic matter, whose energy-momentum tensor violates the null energy condition (NEC).  Recently, in order to circumvent this inconvenience to
get T-WHs, several papers discussed the existence of T-WHs in alternative theories.  Remarkable solutions come from considering one of the most effective generalization of GR to higher dimensions (motivated by the heterotic string theory), known as Einstein-Gauss-Bonnet (EGB) theory. The action for $(n+1)$-dimensional EGB, is 
%%%%
\begin{equation}\label{actionGB}
S[g_{ab},\phi, \psi_{a}] = \int d^{(n+1)}x \sqrt{-g} \left\{ \frac{1}{16\pi}\left(R - \alpha_{_{GB}} R_{_{GB}}^{2} \right) + \frac{1}{4\pi}\mathcal{L}_{\rm matter}(g_{ab},\psi_{a}) \right\},
\end{equation}
where $R_{_{GB}}^{2}$ stands for the quadratic Gauss-Bonnet (GB) term defined by $R_{_{GB}}^{2} = R_{\alpha\beta\mu\nu}R^{\alpha\beta\mu\nu} - 4R_{\alpha\beta}R^{\alpha\beta} + R^{2}$; $\alpha_{_{GB}}$ is known as the GB coefficient (also called  second order Lovelock coefficient), that is proportional to the inverse string tension  \cite{Boulware}. Then assuming that the tension of a string is large compared
to the energy scale, the GB term appears as the first curvature correction to GR \cite{string_con}, \cite{GB_con}. The last term in the action, $\mathcal{L}_{\rm matter}$, is the matter fields Lagrangian and $\psi_{a}$ describes any matter field included in the theory. In this context, for $(n+1)\geq 5$, Lorentzian WH solutions were investigated in \cite{WH_5EGB}. 
%%%%%%%%%%%%%%%%%%%%%%%%%%%%%%%%%%%%%%
It is worth to mention that for  $(n+1)=4$, the GB term does not modify the GR field equations. However if a massless scalar field, $\phi$, is non-minimally coupled to the GB term, through the coupling function $\boldsymbol{f}(\phi)$, then the new field equations are substantially different from the standard GR. In this way  arises one of the best modify-GR theories (in the sector of higher-curvature),  known as the EsGB theory, defined by the action  

\begin{equation}\label{actionL}
S[g_{ab}, \phi, \psi_{a}] = \int d^{4}x \sqrt{-g} \left\{ \frac{1}{16\pi}\left(R - \frac{1}{2}\partial_{\mu}\phi\partial^{\mu}\phi + \boldsymbol{f}(\phi) R_{_{GB}}^{2} \right) + \frac{1}{4\pi}\mathcal{L}_{\rm matter}(g_{ab},\psi_{a}) \right\}.
\end{equation} 
%%%%
This theory arose inspired by an effective string-theory model and its predictions dramatically differ from GR in high-curvature regions, such as the interior of black holes (BHs) and the early universe epochs, where it aims at resolving curvature singularities, and in fact, free curvature singularity BHs, that evade the Penrose singularity theorem \cite{Penrose}, have been numerically derived in EsGB, see \cite{Kanti2018}.  
%%%%%%%%%%%%%%%%%%%%%%
It is also noticeable that the higher curvature GB terms allow the presence of regions with negative effective energy densities \cite{Kunz2011}, and one of the consequences is that structures like T-WHs can be sustained without invoking exotic matter.
In \cite{Kanti2011} linearly stable T-WHs were numerically derived with a coupling function given by $\boldsymbol{f}(\phi)= e^{- \gamma \phi}$ with no need  of exotic matter. 

In this paper, motivated by the recent work in \cite{MagWH}, where by means of magnetic metamaterials and metasurfaces a WH, only for the propagation of the magnetic field, was constructed and experimentally demonstrated, % (this wormhole transfers the magnetic field from one point in space to another through a path that is magnetically undetectable).  % 
we have derived a new T-WH exact solution in EsGB, with a real scalar field (with a positive kinetic term), coupled to a power-Maxwell NLED, $\mathcal{L}(F)= (- \kappa F)^{3/2}$, \cite{Hassaine2008}, \cite{Gurtug2012}. We as well explore the situation of having this T-WH opened by non exotic matter, but rather by the spacetime curvature from the sGB term.
It is fair to mention that since most of the up to now found solutions of the EsGB theory are numerical, it is noteworthy to have derived an exact solution of the EsGB theory coupled to a NLED source.

The paper is organized as follows: in the next section we briefly outline the field equations derived from the EsGB-NLED action. In Sect. \ref{III} we examine general features of wormholes as well as the relation between traversability and the violation of the null energy condition. In Sect. \ref{IV} the derived wormhole solution is presented and the curvature quantities and  null trajectories are analyzed.  In Sect. \ref{V} are treated the limiting cases of vanishing electromagnetic field, and vanishing of the scalar field. Final conclusions are given in the last section.

%{\bf EsGB-NLED system}.
\section{ Einstein-scalar-Gauss-Bonnet theory coupled to nonlinear electromagnetic field  equations}

 Now, we present the dynamical equations of EsGB-NLED. The matter Lagrangian $\mathcal{L}_{\rm matter}$ depends on the electromagnetic invariant $F=\frac{1}{4}F_{ab}F^{ab}$, being  $F_{ab}$ the electromagnetic field tensor defined in terms of the electromagnetic potential $A_{a}$ by $F_{ab} = \partial_{a} A_{b} - \partial_{b} A_{a}$.  We are using units in which $c = G = 1$. Varying the action (\ref{actionL}) with respect to the metric, $g_{ab}$,  the field equations are obtained,
$G_{a}{}^{b} = 8\pi (E_{a}{}^{b})\!_{_{_{S\!G\!B}}} + 8\pi (E_{a}{}^{b})\!_{_{_{N \! L \! E \! D}}}$,
being $(E_{a}{}^{b})\!_{_{_{S\!G\!B}}}$ 
an effective energy-momentum tensor for the sGB term, while $(E_{a}{}^{b})\!_{_{_{N \! L \! E \! D}}}$ corresponds to the NLED energy-momentum tensor, given, respectively, by 
%%%%%%%%%%%%
\begin{eqnarray}
&&8\pi (E_{\alpha}{}^{\beta})\!_{_{_{S\!G\!B}}} = -\frac{1}{4}(\partial_{\mu}\phi \partial^{\mu}\phi)\delta_{\alpha}{}^{\beta} + \frac{1}{2}\partial_{\alpha}\phi \partial^{\beta}\phi - \frac{1}{2}( g_{\alpha\rho} \delta_{\lambda}{}^{\beta} + g_{\alpha\lambda} \delta_{\rho}{}^{\beta})\eta^{\mu\lambda\nu\sigma}\tilde{R}^{\rho\gamma}{}_{\nu\sigma}\nabla_{\gamma}\partial_{\mu}\boldsymbol{f}(\phi), \label{E_GB}\\ 
&&8\pi (E_{\alpha}{}^{\beta})\!_{_{_{N \! L \! E \! D}}} =  2\left(\mathcal{L}\delta_{\alpha}{}^{\beta} -  \mathcal{L}_{F}F_{\alpha\mu}F^{\beta\mu}\right), \label{E_NLED}
\end{eqnarray}
with $\tilde{R}^{\rho\gamma}{}_{\mu\nu} = \eta^{\rho\gamma\sigma\tau}R_{\sigma\tau\mu\nu} = \epsilon^{\rho\gamma\sigma\tau}R_{\sigma\tau\mu\nu}/\sqrt{-g}$, and where  we are denoting the derivative of $\mathcal{L}$ with respect to $F$ by $\mathcal{L}_{F}$. Note that $(E_{a}{}^{b})\!_{_{_{S\!G\!B}}}$ consists on two purely scalar field terms plus a third term related to the coupling between $\phi$ and the GB curvature. Moreover if the coupling function, $\boldsymbol{f}(\phi)$, is  constant, then the tensor  $(E_{\alpha}{}^{\beta})\!_{_{_{S\!G\!B}}}$  reduces to the standard energy-momentum tensor of a massless scalar field. Turning to the field equations, the variation with respect to the electromagnetic potential yields the electromagnetic field equations
$\nabla_{\alpha}( \mathcal{L}_{F} F^{\alpha\beta} ) = 0,$ whereas the variation with respect to $\phi$  yields the motion equation for the scalar field, $\nabla^{2}\phi + \dot{\boldsymbol{f}}(\phi) R_{_{GB}}^{2} = 0,$ where $\dot{\boldsymbol{f}}(\phi)$  denotes the derivative of $\boldsymbol{f}(\phi)$ with respect to the scalar field.

Our aim is to find solutions of EsGB-NLED field equations for static, spherically symmetric, asymptotically flat spacetimes (SSS-AF). Thus, the symmetries of the system imply that the corresponding scalar field must be a function only on $r$, $\phi = \phi(r)$, whereas the metric has the form $ds^{2} =  - e^{ A(r) }dt^{2} + e^{ B(r) }dr^{2}  + r^{2}(d\theta^{2}  + \sin^{2}\theta d\varphi^{2}),$  with $A = A(r)$ and $B = B(r)$ being the metric components to be determined from the EsGB-NLED field equations.

Regarding the electromagnetic field tensor, in a SSS spacetime we can restrict ourselves to purely electric field, then the electromagnetic field tensor has the form, $F_{\alpha\beta} = \mathcal{E}(r)\left( \delta^{t}_{\alpha}\delta^{r}_{\beta} - \delta^{r}_{\alpha}\delta^{t}_{\beta} \right).$ With these assumptions the general solution of the electromagnetic field equations is given by,
%%%%%%%%%%%%%%%%%%%%%%%%%%%%
\begin{equation}\label{fabSOL}
%\partial_{r}( \sqrt{-g} \mathcal{L}_{F} F^{rt}) = 0  \quad \Rightarrow  \quad 
F^{\alpha\beta} = \frac{ \mathcal{Q} }{ \sqrt{- r^{4} g_{tt}g_{rr} } \mathcal{L}_{F} }\delta_{r}{}^{\alpha}\delta_{t}{}^{\beta}  \quad \Rightarrow \quad F = - \frac{ \mathcal{Q}^{2} }{ 2r^{4} \mathcal{L}_{F}^{2}},  
\end{equation}
where $\mathcal{Q}$ is an integration constant that plays the role of the electric charge.  Then the electric field is determined once the Lagrangian is known, as, $\mathcal{E}(r)= \mathcal{Q}  \sqrt{-g_{tt} g_{rr}}/(r^2\mathcal{L}_{F} )$. The explicit field equations can be  consulted in the Appendix.
%%%%%%%%%%%%%%%%%%%%%%%%%%%%%%%%%%%%%%%%%%%%%%

%{\bf Basics on T-WHs.} 
\section{Basics on wormholes and traversability}\label{III}

The canonical metric of a (3+1)-dimensional SSS-WH solution is given by \cite{ThorneMorris}
%%%%%%%%%%%%%%%%%%%%%%%%%%%%%%%%%%%%%%%%%
\begin{equation}\label{WH_Thorne}
ds^{2} =  -  e^{2 \Phi(r)} dt^{2} + \frac{dr^{2}}{ 1  - \frac{b(r)}{r} }  + r^{2} \left( d\theta^{2} + \sin^{2}\theta d\varphi^{2}\right) ,
\end{equation}
where $\Phi(r)$ is called the red-shift function and $b(r)$ is the shape function. The radial coordinate has a range that goes from a minimum value at $r_{0}$, the WH throat defined by $b(r_{0}) = r_{0}$, up to infinity; this coordinate has a special geometric significance, where $4\pi r^2$  is the area of a sphere centered on the WH throat. %%%%%%%%%%%%%%%%%%%%%%%%%%%%%%%%%%%%%%%%%%%%%%%
On the other hand, for the WH to be traversable, one must demand the absence of event horizons, which are identified as the surfaces where $e^{2\Phi(r)} \rightarrow 0$.  If $\Phi(r)$ is a continuous, non vanishing and finite function in the whole range of $r$,  $r\in[r_{0},\infty)$, there will not occur horizons. Moreover, a fundamental property of T-WHs is the fulfillment of the flaring out condition, which is deduced from the mathematics of embedding, and is given by $(b - rb')/b^{2} > 0$, where prime denotes the derivative with respect to $r$. Note that at the throat the flaring out condition reduces to $b'(r_{0})<1$. The condition $(1 - b/r) \geq 0$ is also imposed for all $r\in[r_{0},\infty)$. %values of $r$. 
%%%%%%%%%%%%%%%%%%%%%%%%%%%%%%%%%%%%%%%%%%%%%%%%%%%%%%%%%
%{\bf Traversability and Null Energy Condition.} 
Let us consider a null vector, in the spacetime (\ref{WH_Thorne}),  given by $\boldsymbol{n} = ( e^{-\Phi(r)}, \sqrt{ 1  - b(r)/r }, 0, 0 )$. Then, by identifying $e^{A(r)} = e^{2\Phi(r)}$ and $e^{B(r)} = (1 - b(r)/r)^{-1}$, using (\ref{GttyGrr}) given in the Appendix, and assuming that the flaring out condition is fulfilled, then evaluating the projection of the Einstein tensor on $\boldsymbol{n}$, at $r=r_{0}$ , yields, %%%%%%%%%%%%%
\begin{eqnarray}\label{Gnnr0}
 G_{\alpha\beta}n^{\alpha}n^{\beta}  \Big|_{r=r_{0}} =  -\frac{1}{r^{3}_{0}} \left[ b( r_{0}) - r_{0}b'(r_{0})  \right] =\frac{1}{r^{2}_{0}} \left[ b'
(r_{0})- 1  \right]< 0. 
\end{eqnarray}
%%%%%%%%%%%%%%%%%%%%%%%%%%%%%%%%%%%%%%%%%%%%%%%%%%%%%%%%%
Therefore, in GR  from the balance between the matter and the curvature quantities, $G_{\alpha\beta} = \kappa T_{\alpha\beta}$, the fulfillment of the flaring out condition implies that the NEC (which states that $T_{\alpha\beta}n^{\alpha}n^{\beta}\geq 0$ for any null vector $n^{\alpha}$), is violated; i.e. in GR, the presence of exotic matter is unavoidable for having a T-WH.  
%%%%%%%%%%%%%%%%%
%%%%%%%%%%%%%%%%%%%%%%%%%%%%%%%%%%%%% 
However, in the EsGB-NLED theory, despite the similarity with the GR equations, $G_{\alpha\beta} = 8\pi E^{(e\!f\!f)}_{\alpha\beta}$, with
$E^{(e\!f\!f)}_{\alpha\beta} = (E_{\alpha\beta})\!_{_{_{S\!G\!B}}} + (E_{\alpha\beta})\!_{_{_{N \! L \! E \! D}}}$,  the fulfillment of the flaring out condition does not necessarily demand exotic matter. Using the Eqs. (\ref{Gnnr0}), (\ref{ttyrr}), and the self-consistence of EsGB-NLED, yields
%%%%%%%%%%%%%%%%%%%%%%%%%%
\begin{equation}\label{E_GB_null_vio}
E^{(e\!f\!f)}_{\alpha\beta} n^{\alpha}n^{\beta}\!\Big|_{r=r_{0}} = (E_{\alpha\beta})\!_{_{_{S\!G\!B}}} n^{\alpha}n^{\beta}\!\Big|_{r=r_{0}} = \left[ (E_{r}{}^{r})\!_{_{_{S\!G\!B}}} - (E_{t}{}^{t})\!_{_{_{S\!G\!B}}} \right]\!\Big|_{r=r_{0}} < 0.   
\end{equation}
%%%%%%%%%%%%%%%
Therefore, in a SSS $(3+1)$-spacetime determined by EsGB-NLED, if the violation of NEC happens, the only responsible for is the sGB term. 
%%%%%%%%%%%%%%%%%%%%%%%%%%%%%%%%%%%%%%%%%%%%%%%%%%%%%%%%

%%%%%%%%%%%%%%%%%%%%%%%%%%%%%%%%%%%%%%%%%%%%%%%%%%%%%%%%
%{\bf Novel exact T-WH solution.} 
\section{ Novel exact traversable charged wormhole solution in Einstein-scalar-Gauss-Bonnet coupled to a power-Maxwell NLED }\label{IV}

The  EsGB-power-Maxwel field equations are fulfilled by the following metric,
%%%%%%%%%%%%
\begin{equation}\label{NewSolutionmq}
ds^{2} =  - dt^{2} + \left( 1 - \frac{2 \mathcal{Q}\!_{\rm e}}{r} + \frac{ \mathcal{Q}\!_{_{ \mathcal{S} }}}{r^{2}} \right)^{-1} dr^{2} + r^{2}(d\theta^{2}  + \sin^{2}\theta d\varphi^{2}),
\end{equation}
with %Lagrangian $\mathcal{L}$,
invariant $F$, electric field $\mathcal{E}$, and $\phi$, given respectively by, 
%%%%%%%%%%%%%%%%
\begin{equation}\label{NLED}
%\mathcal{L} =  \left( - \tilde{\kappa} F \right)^{3/2},
%\quad 
F \!=\! - \frac{ (16\mathcal{Q}^2\!_{\rm e})^{\frac{1}{3}} }{ 4\tilde{\kappa} r^{2} },   
\quad \mathcal{E} \!=\! \frac{ (  4 \mathcal{Q}\!_{\rm e}  )^{\frac{1}{3}}  }{ \sqrt{2 \tilde{\kappa} } } \left( r^2 - 2\mathcal{Q}\!_{\rm e} r + \mathcal{Q}_{_{ \mathcal{S} }} \right)^{-\frac{1}{2}}, \quad \phi  \!=\!-\! 2 \ln\!\left( \left|\mathcal{Q}_{_{ \mathcal{S} }} - \mathcal{Q}\!_{\rm e}  r + \sqrt{ \mathcal{Q}_{_{ \mathcal{S} }}\left( r^{2} - 2 \mathcal{Q}\!_{\rm e} r + \mathcal{Q}_{_{ \mathcal{S} }} \right] } \right|/r \right)\!, 
\end{equation}
Using  (\ref{fabSOL}), with $\mathcal{L} =  \left( - \tilde{\kappa} F \right)^{3/2}$, we obtain $F = \sqrt{2}\mathcal{Q} / (3 \tilde{\kappa}^{\frac{3}{2}} r^{2} )$, and by comparison with  (\ref{NLED}), yields $\mathcal{Q} =  - 3 \left(\frac{\mathcal{Q}\!_{\rm e}}{2} \right)^{\frac{2}{3}}\sqrt{ \frac{ \tilde{\kappa} }{ 2}}.$ While the coupling function, $\boldsymbol{f}(\phi(r)) = \int^{r}_{r_{0}}  \phi'(\xi) \dot{\boldsymbol{f}}(\phi(\xi)) d\xi$, is such that 
\begin{equation}\label{f_punto}
\dot{\boldsymbol{f}}(\phi) = \frac{ \mathcal{Q}\!_{\rm e} r^{4} \ln \left( r -\mathcal{Q}\!_{\rm e}  + \sqrt{r^{2} - 2\mathcal{Q}\!_{\rm e}  r + \mathcal{Q}_{_{ \mathcal{S} }} } \right)  }{ 4 \sqrt{\mathcal{Q}_{_{ \mathcal{S} }}} \left( \mathcal{Q}_{_{ \mathcal{S} }} - 2\mathcal{Q}\!_{\rm e}  r\right) }.
\end{equation}  

The line element (\ref{NewSolutionmq}) has  the structure of a SSS-AF-WH metric, with $\Phi=$const. and $b(r)= (2\mathcal{Q}\!_{\rm e}  - \mathcal{Q}\!_{_{ \mathcal{S} }}/r)$; being the WH throat at $r=r_{0} = \mathcal{Q}\!_{\rm e}  + \sqrt{\mathcal{Q}^2\!_{\rm e}  - \mathcal{Q}\!_{_{ \mathcal{S} }} }$, with $\mathcal{Q}^2\!_{\rm e}  > \mathcal{Q}\!_{_{ \mathcal{S} }}$ in order that $r_{0}$ be real. In the wormhole domain, $r \in [r_{0} , \infty)$, the  scalar field is  real and its derivative with respect to $r$, yields
$\phi' = \frac{ 2\sqrt{ \mathcal{Q}_{_{ \mathcal{S} }} } }{ r \sqrt{ r^{2} - 2\mathcal{Q}\!_{\rm e} r + \mathcal{Q}\!_{_{ \mathcal{S} }}  }}$ that diverges at the WH throat. However, this is not a serious problem since $\phi'$ is not a scalar field invariant. In fact,  we can build the invariant quantity $\nabla_{\alpha}\phi \nabla^{\alpha}\phi = g^{rr}\phi'^{2} = \frac{ 4 \mathcal{Q}\!_{_{ \mathcal{S} }} }{ r^{4} },$ that is regular in the whole WH domain.
%%%%%%%%%%%%%%%%%%%%%5 
Regarding the regularity of the WH spacetime, although we know that an actual prove of regularity is the completeness of geodesics,
a way of establishing that the spacetime is non-singular in the WH domain, $r\in[ r_{0} , \infty)$, is evaluating the curvature invariants $R$, $R_{\alpha\beta}R^{\alpha\beta}$ and $R_{\alpha\beta\mu\nu}R^{\alpha\beta\mu\nu}$ and determining if they behave well in the region of interest; for the metric (\ref{NewSolutionmq}) these are,  
%%%%%%%%%%%%%%%%%%%%%%%%%
\begin{eqnarray}
R = \frac{ 2\mathcal{Q}\!_{_{ \mathcal{S} }} }{ r^{4} }, \quad R_{\alpha\beta}R^{\alpha\beta} = \frac{ 2( 3 \mathcal{Q}^2\!_{\rm e}  r^{2} - 4\mathcal{Q}\!_{\rm e}  \mathcal{Q}\!_{_{ \mathcal{S} }} r + 2 \mathcal{Q}^{2}\!\!\!\!_{_{ \mathcal{S} }} ) }{ r^{8} }, \quad R_{\alpha\beta\mu\nu}R^{\alpha\beta\mu\nu} = \frac{ 4( 6 \mathcal{Q}^2\!_{\rm e}  r^{2} - 8\mathcal{Q}\!_{\rm e}  \mathcal{Q}\!_{_{ \mathcal{S} }} r + 3\mathcal{Q}^{2}\!\!\!\!_{_{ \mathcal{S} }} ) }{ r^{8} },
\end{eqnarray}
that are finite in the WH domain. In relation to the NEC, we can see that 
%%%%%%%%%%%%%%%%%
\begin{equation}
\frac{b - rb'}{b^{2}} = \frac{2( \mathcal{Q}\!_{\rm e} r - \mathcal{Q}\!_{_{ \mathcal{S} }} ) r }{( 2\mathcal{Q}\!_{\rm e} r - \mathcal{Q}\!_{_{ \mathcal{S} }} )^{2} } > 0 \quad \textup{and} \quad b'(r_{0}) = \frac{  \mathcal{Q}\!_{_{ \mathcal{S} }} }{\left( \mathcal{Q}\!_{\rm e}  + \sqrt{\mathcal{Q}^2\!_{\rm e}  - \mathcal{Q}\!_{_{ \mathcal{S} }} }\right)^{2} }<1,
\end{equation}
%%%%%%%%%%%%%%%
then, the solution (\ref{NewSolutionmq}) admits a T-WH interpretation. 
%%%%%%%%%%%%%
Moreover, in agreement with (\ref{E_GB_null_vio}) we get,
%%%%%%%%% 
%%%%%%%
\begin{equation}
8\pi (E_{\alpha\beta})\!_{_{_{S\!G\!B}}}n^{\alpha}n^{\beta}\Big|_{r=r_{0}} = - \frac{2( \mathcal{Q}^2\!_{\rm e} - \mathcal{Q}\!_{_{ \mathcal{S} }} + \mathcal{Q}\!_{\rm e}  \sqrt{ \mathcal{Q}^2\!_{\rm e}  - \mathcal{Q}\!_{_{ \mathcal{S} }} }  )}{ (\mathcal{Q}\!_{\rm e}  + \sqrt{ \mathcal{Q}^2\!_{\rm e}  - \mathcal{Q}\!_{_{ \mathcal{S} }} }  )^{4} }   < 0,
\end{equation}
%%%%%%
that is negative since $\mathcal{Q}^2\!_{\rm e}  > \mathcal{Q}\!_{_{ \mathcal{S} }}$. Thus, we have proved that the solution (\ref{NewSolutionmq}), with $\mathcal{Q}^2\!_{\rm e}  > \mathcal{Q}\!_{_{ \mathcal{S} }}$, has a T-WH interpretation.

\begin{figure}
\centering
\includegraphics[width=17.5cm,height=6cm]{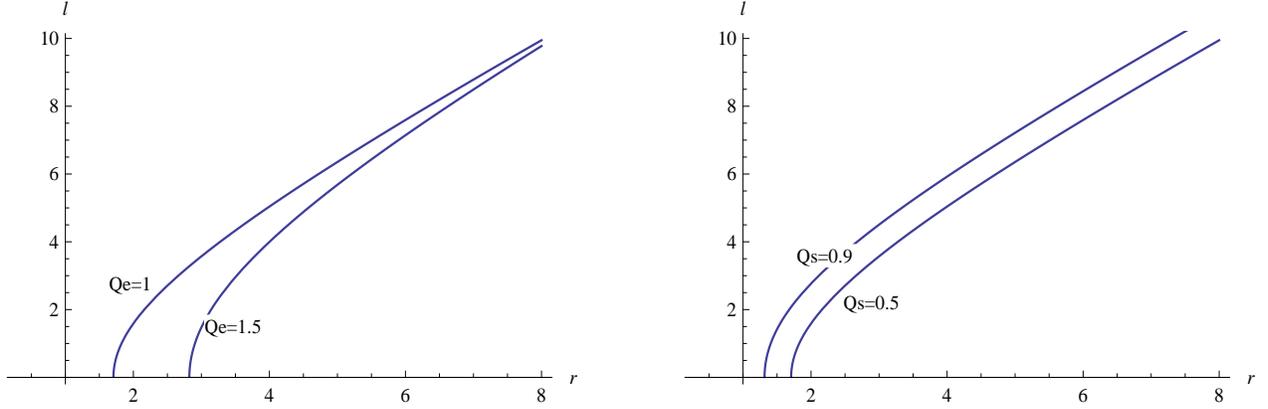}
\caption{\label{fig1} The proper radial distance (the positive $l(r)$ ) for the T-WH is illustrated; the throat is located at $l(r) = 0$. To the left is shown for two different $\mathcal{Q}\!_{\rm e} =1; 1.5$ (the scalar charge $\mathcal{Q}_{_{ \mathcal{S} }}=0.5$), the effect of increasing $\mathcal{Q}\!_{\rm e} $ is to enhance the throat.  To the right it is shown $l(r)$ for two values of the scalar charge $\mathcal{Q}_{_{ \mathcal{S} }}= 0.5;0.9$, in this case the effect is the opposite, increasing the scalar charge shrinks the throat (in this plot $\mathcal{Q}\!_{\rm e} =1$).}
\end{figure}

%%%%%%%%%%%%%%%%%%%%%%%%%%%%%%%%%%%%%%%%%%%%%%%%%%%
In the general case, the radial coordinate $r$ is ill behaved near the throat; however, the proper radial distance $l(r)$, defined as $l(r)= \pm \int_{r_0}^{r}{\frac{ d\tilde{r} }{\sqrt{1-\frac{b(\tilde{r})}{\tilde{r}}}}},$ 
must be well behaved everywhere. %; it is required that $l(r)$ be finite throughout spacetime. %, which also implies that $1-{b(r)}/{r} \ge 0$. 
The plus (minus) sign is related to the upper (lower) part of the wormhole. For (\ref{NewSolutionmq}), 
\begin{equation}
 l(r) = \pm\left(\!\sqrt{r^2\!-\!2\mathcal{Q}\!_{\rm e} r \!+\! \mathcal{Q}\!_{_{ \mathcal{S} }} } \!+\! \mathcal{Q}\!_{\rm e}  \ln{\!\!\left[r \!-\! \mathcal{Q}\!_{\rm e} \! +\! \sqrt{ r^2 \!-\! 2\mathcal{Q}\!_{\rm e} r \!+\! \mathcal{Q}\!_{_{ \mathcal{S} }} }\right]} \!-\! \mathcal{Q}\!_{\rm e} \ln{\!\!\left[\sqrt{ \mathcal{Q}^2\!_{\rm e} \! -\! \mathcal{Q}\!_{_{ \mathcal{S} }} } \right]}  \right).
\end{equation}
The effect of varying the WH parameters on the proper radial distance is illustrated in Fig. \ref{fig1}.
%%%%%%%%%%%%%%%%%%%%%%%%%%
%***********************fig1*proper*radial*distance*********
%%%%%%%%%%%%%%%%%%%%%%%%%%%%%%%%%%%%%%%%%%%%%%%%%%%%%%%

{\bf Photon trajectories and capture cross-section for light.} In order to determine the capture cross-section for photons in the neighborhood of the WH as well as the orbits around the WH throat,   the metric is written in more suitable coordinates.  With $r = \sqrt{ \rho^{2} + r_{0}^{2} }$,  (\ref{NewSolutionmq}) takes the form,
%%%%%%%%%%%%%%%%%%%%%%%
\begin{equation}\label{genr_Ellis}
ds^{2} =  - dt^{2} + \frac{ \rho^{2}  }{\left( \sqrt{ \rho^{2} + r_{0}^{2} } - r_{0} \right) \left( \sqrt{ \rho^{2} + r_{0}^{2} } - r_{1} \right)  } d\rho^{2}  + (\rho^{2} + r_{0}^{2} )(d\theta^{2}  + \sin^{2}\theta d\varphi^{2}), 
\end{equation}
with $r_{0} =  \mathcal{Q}\!_{\rm e}  + \sqrt{  \mathcal{Q}^2\!_{\rm e}  - \mathcal{Q}\!_{_{ \mathcal{S} }}  }$,   $r_{1} =  \mathcal{Q}\!_{\rm e}  - \sqrt{  \mathcal{Q}^2\!_{\rm e}  - \mathcal{Q}\!_{_{ \mathcal{S} }}  }$    and  $\mathcal{Q}^2\!_{\rm e}  > \mathcal{Q}\!_{_{ \mathcal{S} }}.$
The new radial coordinate $\rho$ has domain $-\infty < \rho < \infty$, thus the limits $\rho \rightarrow \pm \infty$ correspond to two distinct asymptotically flat regions, whereas the WH throat is at $\rho =0$. 
On the other hand, the photon trajectories are described by the equations, 
%%%%%%%%%%%%%%%%%%%
\begin{equation}\label{nullgeo}
\frac{d^{2}x^{\alpha}}{d\lambda^{2}} + \Gamma^{\alpha}_{\beta\nu} \frac{dx^{\beta}}{d\lambda}\frac{dx^{\nu}}{d\lambda} = 0, \quad\quad \textup{ and } \quad\quad ds^{2}(u,u) = 0.   
\end{equation}
where $\lambda$ and $u = \frac{ dx^{\alpha} }{ d\lambda }\partial_{ \alpha } $ are the affine parameter and the
null vector of  null geodesics, respectively.
%%%
%%%
Symmetries imply the conservation of the energy, $\mathsf{E},$ and angular momentums, $\ell$ and  $\mathsf{L}_{\theta},$ for the test particle, $\mathsf{E} = \frac{dt}{d\lambda}$, $\ell = \left( \rho^{2} + r_{0}^{2} \right)\frac{d\varphi}{d\lambda}$ and $\mathsf{L}_{\theta} = \left( \rho^{2} + r_{0}^{2} \right)\frac{d\theta}{d\lambda}.$
%%%
%%%
Conservation of the angular momentum, $\ell$, or of $\mathsf{L}_{\theta}$, means that the particle will move on a plane. Thus, without loss of generality, one can choose the plane $\theta = \frac{\pi}{2}$.
Eq. (\ref{nullgeo}) for the $\rho$-component amounts to
%%%
\begin{equation}
\left( \frac{d\rho}{d\lambda} \right)^{2} + \left[ \left( \mathsf{E}^{2} - \frac{\ell^{2}}{ \rho^{2} + r_{0}^{2} }  \right)\frac{  (r_{0} + r_{1})\sqrt{  \rho^{2} + r_{0}^{2}  } - r_{0}^{2} - r_{0}r_{1}  }{ \rho^{2} }  + \frac{\ell^{2}}{ \rho^{2} + r_{0}^{2} } \right] = \mathsf{E}^{2},  
\end{equation}
%%%%%%%%%%%%%
which can be written as $\left( \frac{d\rho}{d\lambda} \right)^{2} + \mathsf{V}_{\!e\!f\!f}^{2}(\rho) = \mathsf{E}^{2}$, with the effective potential given by
%%%%
\begin{equation}
\mathsf{V}_{\!e\!f\!f}^{2}(\rho) = \left[ \left( \mathsf{E}^{2} - \frac{ \ell^{2} }{ \rho^{2} + r_{0}^{2} }  \right)\frac{  ( r_{0} + r_{1} )\sqrt{  \rho^{2} + r_{0}^{2}  } - r_{0}^{2} - r_{0}r_{1}  }{ \rho^{2} }  + \frac{ \ell^{2} }{ \rho^{2} + r_{0}^{2} } \right].
\end{equation}
%%%%
 
The potential is maximum at $\rho = 0$, ($\mathsf{V}_{\!e\!f\!f}^{Max} = \mathsf{E}$), and this corresponds to an unstable circular orbit. For energies ($\mathsf{E} > \mathsf{V}_{\!e\!f\!f}^{Max}$) the photon goes through the wormhole throat, and comes out on the other side of the throat. In contrast, for energies ($\mathsf{E} < \mathsf{V}_{\!e\!f\!f}^{Max}$) an incoming photon is scattered  to infinity.   
%***********************fig2*effective*potential***********
\begin{figure}
\centering
\includegraphics[width=17.5cm,height=6cm]{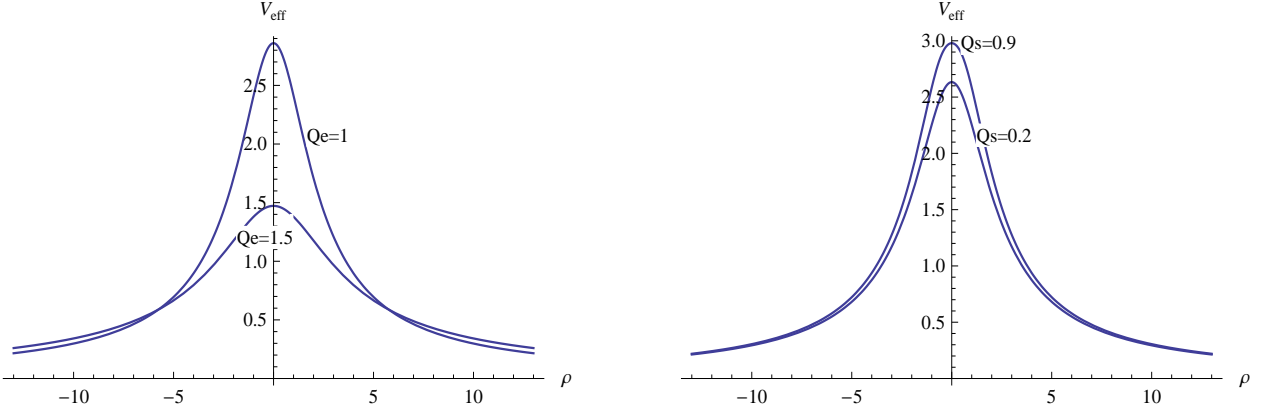}
\caption{\label{fig2} The effective potential as a function of $\rho$ for a massless test particle, in the T-WH geometry, is illustrated. To the left is shown for two different $\mathcal{Q}\!_{\rm e} =1; 1.5$ (the impact parameter $b=4$ and the scalar charge $\mathcal{Q}_{_{ \mathcal{S} }}=0.5$), the effect of increasing the electromagnetic charge $\mathcal{Q}\!_{\rm e}$ is to lower the potential barrier. To the right it is shown the effective potential for two values of the scalar charge $\mathcal{Q}_{_{ \mathcal{S} }}= 0.2;0.9$; for a greater scalar charge the potential barrier is higher (the impact parameter $b=4$ and  $\mathcal{Q}\!_{\rm e} =1$).}
\end{figure}
%%%
%***************************************************************
The effective potential $\mathsf{V}_{\!e\!f\!f}^{Max}$  for a photon in the unstable circular orbit, is  
%%%%%%%%%%%%
\begin{equation}
\left( \mathsf{V}_{\!e\!f\!f}^{Max} \right)^{2} = \lim\limits_{\rho \to 0} \mathsf{V}_{\!e\!f\!f}^{2}(\rho) =  \frac{( r_{0} + r_{1})\left( \mathsf{V}_{\!e\!f\!f}^{Max} \right)^{2}}{2r_{0}} + \frac{( r_{0} - r_{1} )\ell^{2}}{ 2r_{0}^{3} },
\end{equation} 
which yields $\mathsf{V}_{\!e\!f\!f}^{Max} = \frac{ \ell }{ r_{0} }$
%%%%%%%%
and then according to \cite{EllisLensing}, the impact parameter $\boldsymbol{\mathrm{b}}$ will be  $\boldsymbol{\mathrm{b}} =  \ell/\mathsf{V}_{\!e\!f\!f}^{Max} = r_{0}$, and hence the capture cross-section for a light beam is $\sigma = \pi\boldsymbol{\mathrm{b}}^{2} = \pi r_{0}^{2}.$ Plots of the effective potential illustrate the above described behavior, see Fig. \ref{fig2}.
%%%%%%%%%%%%%%%%%%%%%%%%%%%%%%%%%%%%%%%%%%%%%%%%%%%%%%

%\section{Limit cases, $\mathcal{M}\!_{_{ \mathcal{E} }}=0$, and $\mathcal{Q}\!_{_{ \mathcal{S} }} = 0$}
%{\bf Uncharged, $\mathcal{Q}\!_{\rm e} =0$, and trivial scalar field, $\mathcal{Q}\!_{_{ \mathcal{S} }} = 0$, cases.} 

\section{ Uncharged, $\mathcal{Q}\!_{\rm e} =0$, and trivial scalar field, $\mathcal{Q}\!_{_{ \mathcal{S} }} = 0$, cases }\label{V}

Below, we will consider two limiting cases of vanishing of one of the charges $\mathcal{Q}\!_{\rm e} =0$, or $\mathcal{Q}\!_{_{ \mathcal{S} }} = 0$. In the case of vanishing electromagnetic field, $\mathcal{Q}\!_{\rm e} = 0$, the scalar field becomes a phantom one keeping the WH open and in this case the well-known Ellis WH solution \cite{Ellis} is recovered. The second case is the vanishing of the scalar field charge, $\mathcal{Q}\!_{_{ \mathcal{S} }} = 0$; in this limit the GB term and the scalar field decouple, but the WH remains traversable, being supported by the GB curvature and the NLED field, with no need of exotic matter.
 
%%%%%%%%%%%%%%%%%%%%%%%%%%%%%%%%%%%%%%%%%%%%%%%%%%%%%
{\bf The Ellis WH.} If $\mathcal{Q}\!_{\rm e} \!=\! 0$, then Eq. (\ref{NLED}) implies that $F\!=\!0\!=\!\mathcal{L}$. Hence, in this case all of the electromagnetic energy-momentum tensor components vanish, i.e., $(E_{\alpha\beta})\!_{_{_{N \! L \! E \! D}}} = 0$.  Moreover, from Eq.  (\ref{NLED}) we find that the scalar field becomes,
%%%%%%%%%%%%%%%%
$\phi = - \ln\left( \frac{ \left( \mathcal{Q}_{_{ \mathcal{S} }} + \sqrt{ \mathcal{Q}_{_{ \mathcal{S} }}\left( r^{2}  + \mathcal{Q}_{_{ \mathcal{S} }} \right) } \right)^{2} }{r^2} \right).$ 
%%%%%%%
In this case, if $\mathcal{Q}\!_{\rm e}  = 0$, then $\dot{\boldsymbol{f}}(\phi) = 0$ implying that $\boldsymbol{f}(\phi)=$const. Therefore the third term in Eq. (\ref{E_GB}) vanishes; then $(E_{\alpha\beta})\!_{_{_{S\!G\!B}}}$ becomes the energy-momentum tensor of a massless scalar field. 
%%%%%%%%%%%%%%%%%%%%%%%%%%%%%%%%%
We will show that in this case the line element (\ref{NewSolutionmq})  still can be interpreted as a WH if the parameter $\mathcal{Q}_{_{ \mathcal{S} }}$ is negative. Setting $\mathcal{Q}_{_{ \mathcal{S} }} = - q^{2}$,  the line element (\ref{NewSolutionmq}), becomes
%%%%%%%%%%%%%%%%%%%%%%%%%%%
\begin{equation}\label{GB_WH_V}
ds^{2} =  - dt^{2} + \left(  1  - \frac{ q^{2} }{r^{2}} \right)^{-1} dr^{2}  + r^{2}(d\theta^{2}  + \sin^{2}\theta d\varphi^{2}), \quad\quad r\in[ |q| , \infty ). 
\end{equation}
Whereas the scalar field takes the form,
%%%
$\phi \!=\! - \ln\left( \frac{ \left( -q^{2} + \sqrt{ -q^{2}\left( r^{2}  - q^{2} \right) } \right)^{2} }{r^2} \right),$
%%%%%%%%%%%%
which can be written as 
\begin{equation}
\phi = \phi_{0} + 2i\tan^{-1}\left( \sqrt{ \frac{r^{2}  - q^{2}}{q^{2}} } \right), \quad\textup{ with }\quad \phi_{0} = -\ln(4q^{2}).
\end{equation}
%%%%
Now, since to the EsGB field equations only contribute $\phi'(r)$ and $\phi''(r)$, we can drop the constant $\phi_{0}$ and consider that the corresponding scalar field is purely imaginary. Then a new scalar field can be defined as $\psi = - i\phi$ (phantom scalar field), and then the line element (\ref{GB_WH_V}) becomes a solution of the action,  
%%%%%%%%%%%%%%%%%%%%%%%%%%%%%%
\begin{equation}\label{actionL2}
S[g_{ab},\psi] = \int d^{4}x \sqrt{-g} \left\{ \frac{1}{16\pi}\left(R + \frac{1}{2}\partial_{\mu}\psi\partial^{\mu}\psi  \right)  \right\},
\end{equation} 
with $\psi$ being a massless phantom scalar field given by,
%%%
$\psi =  2\tan^{-1}\left( \sqrt{ \frac{r^{2}  - q^{2}}{q^{2}} } \right).$
%%%%%
This spacetime fulfills that 
%%%%%%%%
$\frac{b - rb'}{b^{2}} = \frac{2r }{q^{2} } > 0$, $b'(r_{0}) = -1,$  
 and $8\pi E_{\alpha\beta}n^{\alpha}n^{\beta}\Big|_{r=r_{0}} = - \frac{2}{ q^{2} } < 0,$
%%%%%%%
indicating that the flaring out condition is satisfied, and therefore it admits a T-WH interpretation. 
 %%%%%%%%%%%%%
The T-WH, (\ref{GB_WH_V}), was originally reported by H.G. Ellis in \cite{Ellis}. Specifically, in \cite{Ellis} it is shown that (\ref{GB_WH_V}) is a GR solution with a minimally coupled phantom scalar field, i.e, (\ref{actionL2}). Alternatively, in \cite{Ellis_EsGB}, was shown that the Ellis wormhole also can be understood as an electrovacuum solution of EsGB gravity with
a material source consisting of a source-free electric field and a real scalar field.
%%% Alternatively, in \cite{Ellis_EsGB}, was showed that the Ellis wormhole also can understand as an electrovacuum solution of EsGB gravity %%%with a material source consisting of source-free electric field and a real scalar field.

The Ellis WH have been extensively studied, %for instance in 
and its properties like gravitational lensing  \cite{EllisLensing}, quasi-normal modes \cite{QNMEllis}, 
stability \cite{stability} and shadows \cite{EllisShadows}, have been thoroughly investigated.
%%%%%%%%%%%%%%%%%%%%%%%%%
 In this case the metric (\ref{genr_Ellis}) reduces to  
\begin{equation}
ds^{2} =  - dt^{2} + d\rho^{2}  + (\rho^{2} + q^{2})(d\theta^{2}  + \sin^{2}\theta d\varphi^{2}), 
\end{equation}
 that is the Ellis-WH.  Hence the T-WH (\ref{NewSolutionmq}) here presented is a generalization of the Ellis WH.
%%%%%%%%%%%%%%%%%%%%%%%%%%%%%%%%%%%%%%%%%%%%%%%%%%%%%%%%%%

{\bf EsGB-NLED in the trivial scalar field limit.} The trivial scalar field, $\phi(r) =$const,  is obtained by setting $\mathcal{Q}_{_{ \mathcal{S} }} = 0$; in this case the metric (\ref{NewSolutionmq}) reduces to,
%%%%%%%%%%%%%%%%%%%%%%%%%%%%%%%%%%%%%% 
\begin{equation}\label{NewSolutionmq0}
ds^{2} =  - dt^{2} + \left(  1 - \frac{2 \mathcal{Q}\!_{\rm e} }{r}  \right)^{-1} dr^{2}  + r^{2}(d\theta^{2}  + \sin^{2}\theta d\varphi^{2}), \quad r\in[ 2 \mathcal{Q}\!_{\rm e} , \infty) \quad \textup{with} \quad \mathcal{Q}\!_{\rm e} > 0,
\end{equation}
that is a T-WH with domain $r\in  [2\mathcal{Q}\!_{\rm e} , \infty)$, since $(b - rb')/b^{2} = ( 2\mathcal{Q}_{\rm e})^{-1}  > 0$ and $b'(r_{0}) = 0$. $\mathcal{Q}_{_{ \mathcal{S} }} = 0$ also
implies that $\phi'= 0$, $\phi'' = 0$, while $\dot{\boldsymbol{f}}(\phi)$ and $\ddot{\boldsymbol{f}}(\phi)$ diverge. But since the quantities $\dot{\boldsymbol{f}}(\phi)$ and $\ddot{\boldsymbol{f}}(\phi)$ appear in the field equations as   
%%%%%%%%%
\begin{equation}
\phi'\dot{\boldsymbol{f}}(\phi) = - \frac{   r^{2} \ln \left( r - \mathcal{Q}\!_{\rm e}  + \sqrt{ r^{2} - 2\mathcal{Q}\!_{\rm e}  r  } \right)  }{ 4   \sqrt{ r^{2} - 2\mathcal{Q}\!_{\rm e} r   } }, \quad 
 \phi''\dot{\boldsymbol{f}}(\phi) =  \frac{ (3\mathcal{Q}\!_{\rm e} r - 2r^{2})  \ln \left( r - \mathcal{Q}\!_{\rm e}  + \sqrt{ r^{2} - 2\mathcal{Q}\!_{\rm e}  r  } \right)  }{ 4  \left(  2\mathcal{Q}\!_{\rm e}  - r\right) \sqrt{ r^{2} - 2\mathcal{Q}\!_{\rm e} r  } }, 
\end{equation}
and     
%%%%%%%%%%
\begin{equation}
\phi'^{2}\ddot{\boldsymbol{f}}(\phi) = \frac{ r + 3 \sqrt{r^{2} - 2\mathcal{Q}\!_{\rm e} r } \ln \left( r - \mathcal{Q}\!_{\rm e}  + \sqrt{r^{2} - 2\mathcal{Q}\!_{\rm e}  r  } \right)   }{ 4(2\mathcal{Q}\!_{\rm e}  - r )},
\end{equation}
%%%%%%
that is easy to prove they  are well defined in the whole WH domain. Therefore there is no conflict with the field equations when $\mathcal{Q}_{_{ \mathcal{S} }} = 0$.
%%%%%%%%%%%%%%%%%%%%%%%%%%%%%%%%%%%%
In this case, since the scalar field is trivial, then the condition (\ref{E_GB_null_vio}) is fulfilled without any kind of matter needed, and the only responsible for keeping the wormhole throat open is the GB gravity. 
%%%%%%%%%%%%%%%%%%%%%%%%%%%%%%%%%%%%%

%%%%%%%%%%%%%%%%%%%%%%%%%%%%%%%%%%%%%%%%%%%%%%%%%%%%%%%
%{\bf Conclusions.} 
\section{Conclusions}

We have derived a new SSS-AF-T-WH solution to the EsGB-power-Maxwell field equations, for a real scalar field with a positive kinetic term. The parameters of the T-WH solution are such that, increasing the electric charge the throat is enhanced, while increasing the scalar charge it shrinks. The light trajectories agree with these effects, since increasing the electric charge makes lower the effective potential barrier, contrasting with the increasing of the scalar charge that makes higher the barrier; meaning that to larger $\mathcal{Q}\!_{\rm e}$ corresponds a larger capture cross section for light, while to larger $\mathcal{Q}\!_{_{ \mathcal{S} }}$ corresponds a smaller capture cross section for light.
%%%%%%%%%%%
The limiting cases, absence of electromagnetic field, $\mathcal{Q}\!_{\rm e}=0$, and vanishing of scalar charge,  $\mathcal{Q}\!_{_{ \mathcal{S} }}=0$ were analyzed and in both cases the WH preserves its traversability, in the former case it is sustained by a phantom scalar field, while in the latter case the WH is kept open by the GB curvature with no need of exotic matter. 
%%%%%%%%%%%%%%%%%%%%%%%%%%%%%%%%%%%%%%%%%%%%%%%%%%%
In a forthcoming paper it shall be  determined the quasinormal modes, the corresponding Kruskal-Szekeres and Penrose diagrams, as well as we shall explore the stability of this T-WH.
\vspace{0.5cm}

\textbf{Acknowledgments}: N. B. and P. C. acknowledges partial financial support from CONACYT-Mexico through the project No. 284489. P. C. also thanks Cinvestav for hospitality.\\

%%%%%%%%%%%%%%%%%%%%%%%%%%%%%%%%%%%%%%%%%%%%%%%%%%%%%

\section*{Appendix}

In this appendix we include the explicit form of the field equations that are satisfied by the wormhole solution (\ref{NewSolutionmq})-(\ref{NLED})-(\ref{f_punto}). \\ %(\ref{SSSmet}).\\
%%%%%%%%%%%%%%%%%%%%%%%%%%%%%%%%%%%%%%%%%%%%%%%%%%%%%%%%%%% 

The nonvanishing Einstein tensor components %for the line element (\ref{SSSmet}) 
are given by 

\begin{eqnarray}
&& G_{t}{}^{t} = \frac{ e^{ -B} }{ r^{2} }\left( -rB' - e^{ B} + 1 \right), \quad\quad  G_{r}{}^{r} = \frac{ e^{ -B} }{ r^{2} }\left( rA' - e^{ B} + 1 \right), \label{GttyGrr}\\   
&& G_{\phi}{}^{\phi} = G_{\theta}{}^{\theta} = \frac{ e^{ -B} }{ 4r } \left( rA'^{2} - rA'B' + 2rA'' + 2A' - 2B' \right)  
\label{GththyGphph}
\end{eqnarray}
%where prime denotes the derivative with respect to $r$.
%%%%%%%%%%%%%%%%%%%%%%%%%%%%%%%%%%%%%%%%%%%%%%%%%%%%%%%%%%%%%%%%%%

The non-vanishing  sGB-energy-momentum tensor components are,

\begin{eqnarray}
&& 8\pi (E_{t}{}^{t})\!_{_{_{S\!G\! B}}} = -\frac{ e^{ -2B} }{ 4r^{2} }\left\{  \left[r^{2}e^{B} + 16(e^{B} - 1)\ddot{\boldsymbol{f}} \right]\phi'^{2} - 8[ (e^{B} - 3)B'\phi' - 2(e^{B} - 1)\phi'']\dot{\boldsymbol{f}} \right\} \\ \label{GBtt}
&& 8\pi (E_{r}{}^{r})\!_{_{_{S\!G\! B}}} = \frac{ e^{ -B} \phi'}{ 4 } \left[ \phi'  - \frac{8(e^{B} - 3)e^{-B}A'\dot{\boldsymbol{f}} }{r^{2}}  \right]  \label{GBrr}\\   
&& 8\pi (E_{\theta}{}^{\theta})\!_{_{_{S\!G\! B}}} = 8\pi (E_{\phi}{}^{\phi})\!_{_{_{S\!G\! B}}} = -\frac{ e^{ -2B} }{ 4r } \left\{ ( re^{B} - 8A'\ddot{\boldsymbol{f}})\phi'^{2}    - 4\left[ (A'^{2} + 2A'')\phi' + (2\phi'' - 3B'\phi')A'\right]\dot{\boldsymbol{f}} \right\}
\end{eqnarray}
%%%%%%%%%%%%%%%%%%%%%%%%%%%%%%%%%%%%%%%%%%%%%%%%%%%%%%%%%%%%%%%%%%%
Note that $\dot{\boldsymbol{f}}$ and $\ddot{\boldsymbol{f}}$ appear always multiplied by derivatives of $\phi$,
as $(\phi'' \dot{\boldsymbol{f}})$,  $(\phi')^2 \ddot{\boldsymbol{f}}$ and  $(\phi' \dot{\boldsymbol{f}})$. \\
%%%%%%%%%%%%%%%%%%%%%%%%%%%%%%%%%%%%%%%%%%%%%%%%%%%%%%%%%%%%%%%%%%%

The NLED-energy-momentum tensor components are given by 

\begin{eqnarray}
&& 4\pi (E_{t}{}^{t})\!_{_{_{N \! L \! E \! D}}} = 4\pi (E_{r}{}^{r})\!_{_{_{N \! L \! E \! D}}} = \mathcal{L} - 2F\mathcal{L}_{F}   \label{ttyrr}\\   
&& 4\pi (E_{\theta}{}^{\theta})\!_{_{_{N \! L \! E \! D}}} = 4\pi (E_{\phi}{}^{\phi})\!_{_{_{N \! L \! E \! D}}} = \mathcal{L} 
\end{eqnarray}
%%%%%%%%%%%%%%%%%%%%%%%%%%%%%%%%%%%%%%%%%%%%%%%%%%%%%%%%%%% 

With the previous equations we establish the EsGB-NLED field equations for the SSS metric as,

\begin{eqnarray}
&& G_{t}{}^{t} = 8\pi E^{(e\!f\!f)}{}_{t}{}^{t} \quad \Rightarrow \quad   4e^{B}\left( rB' + e^{ B} - 1 \right) =  \left[ r^{2}e^{B} +16(e^{ B}-1)\ddot{\boldsymbol{f}} \right]\phi'^{2} - 8  \left[ (e^{ B} - 3)B'\phi' - 2(e^{ B} - 1)\phi'' \right]\dot{\boldsymbol{f}} \nonumber \\ 
&& \hskip6.6cm + \!\! \quad   8r^{2}e^{2B}(2F\mathcal{L}_{F} - \mathcal{L}),\label{Eqt}\\
&&G_{r}{}^{r} = 8\pi E^{(e\!f\!f)}{}_{r}{}^{r} \quad \Rightarrow \quad   4e^{B}\left( -rA' + e^{ B} - 1 \right) =   -r^{2}e^{B}\phi'^{2} + 8(e^{ B} - 3)A'\phi'\dot{\boldsymbol{f}} + 8r^{2}e^{2B}(2F\mathcal{L}_{F} - \mathcal{L}), \label{Eqr}\\
&&G_{\theta}{}^{\theta} = 8\pi E^{(e\!f\!f)}{}_{\theta}{}^{\theta} \quad \Rightarrow \quad  e^{B}\left\{ rA'^{2} - 2B' + (2 - rB')A'   + 2rA''   \right\} = -re^{B}\phi'^{2} + 8A'\ddot{\boldsymbol{f}}\phi'^{2} \nonumber \\ 
&& \hskip6.6cm + \!\! \quad 4  \left[  (A'^{2} + 2A'')\phi' + (2\phi'' - 3B'\phi')A' \right]\dot{\boldsymbol{f}} +  8re^{2B}\mathcal{L}.\label{Eqte}
\end{eqnarray}
Whereas the scalar field should satisfy

\begin{equation}
2r\phi'' + (4 + rA' - rB')\phi' + \frac{4e^{-B}\dot{\boldsymbol{f}}}{r} \left[ (e^{B} - 3)A'B' - (e^{B} - 1)(2A'' + A'^{2})\right] = 0.   
\end{equation}

\end{document}